\documentclass[twoside,twocolumn,9pt]{article}
\usepackage{amssymb}
\usepackage{amsmath}
\usepackage{bm}
\usepackage{extsizes}
\usepackage[super,sort&compress,comma]{natbib} 
\usepackage[version=3]{mhchem}
\usepackage[left=1.5cm, right=1.5cm, top=1.785cm, bottom=2.0cm]{geometry}
\usepackage{balance}
\usepackage{times,mathptmx}
\usepackage{sectsty}
\usepackage{graphicx} 
\usepackage{lastpage}
\usepackage[format=plain,justification=justified,singlelinecheck=false,font={stretch=1.125,small,sf},labelfont=bf,labelsep=space]{caption}
\usepackage{float}
\usepackage{fancyhdr}
\usepackage{fnpos}
\usepackage[english]{babel}
\usepackage{array}
\usepackage{droidsans}
\usepackage{charter}
\usepackage[T1]{fontenc}
\usepackage[usenames,dvipsnames]{xcolor}
\usepackage{setspace}
\usepackage[compact]{titlesec}
\usepackage{hyperref}

\usepackage{epstopdf}

\usepackage[squaren]{SIunits} 
\usepackage{bbm} 

\newcommand{\bn}{}

\DeclareMathOperator{\sgn}{sgn}

\definecolor{cream}{RGB}{222,217,201}

\begin{document}
\pagestyle{fancy}
\thispagestyle{plain}
\fancypagestyle{plain}{

\fancyhead[C]{\includegraphics[width=18.5cm]{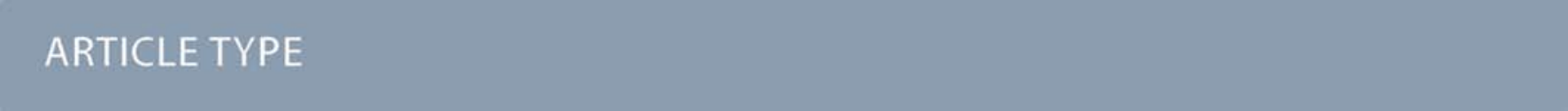}}
\fancyhead[L]{\hspace{0cm}\vspace{1.5cm}\includegraphics[height=30pt]{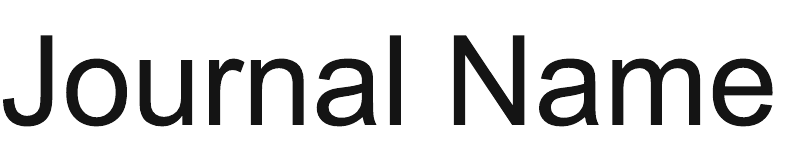}}
\fancyhead[R]{\hspace{0cm}\vspace{1.7cm}\includegraphics[height=55pt]{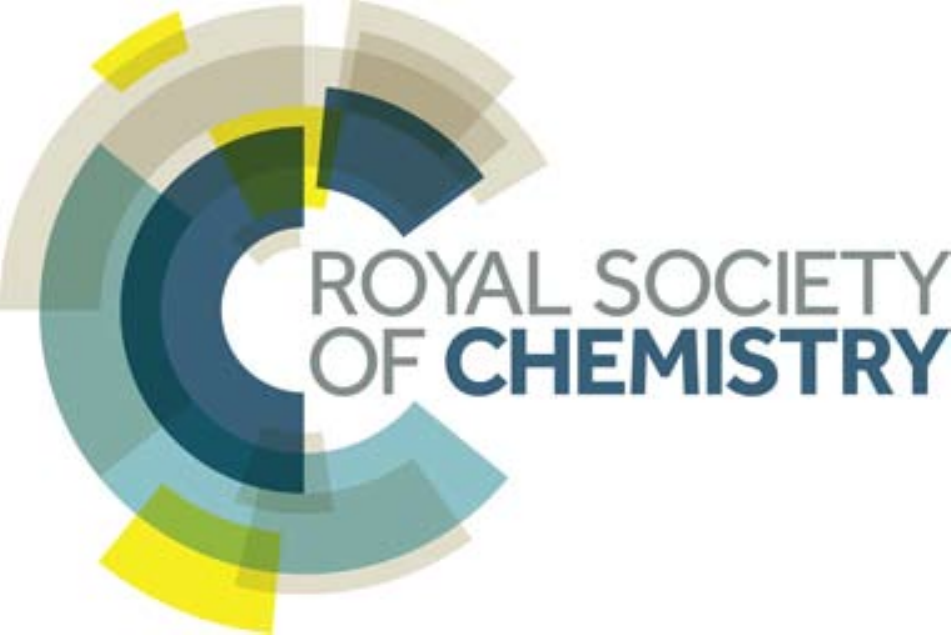}}
\renewcommand{\headrulewidth}{0pt}
}

\makeFNbottom
\makeatletter
\renewcommand\LARGE{\@setfontsize\LARGE{15pt}{17}}
\renewcommand\Large{\@setfontsize\Large{12pt}{14}}
\renewcommand\large{\@setfontsize\large{10pt}{12}}
\renewcommand\footnotesize{\@setfontsize\footnotesize{7pt}{10}}
\makeatother

\renewcommand{\thefootnote}{\fnsymbol{footnote}}
\renewcommand\footnoterule{\vspace*{1pt}%
\color{cream}\hrule width 3.5in height 0.4pt \color{black}\vspace*{5pt}} 
\setcounter{secnumdepth}{5}

\makeatletter 
\renewcommand\@biblabel[1]{#1}            
\renewcommand\@makefntext[1]%
{\noindent\makebox[0pt][r]{\@thefnmark\,}#1}
\makeatother 
\renewcommand{\figurename}{\small{Fig.}~}
\sectionfont{\sffamily\Large}
\subsectionfont{\normalsize}
\subsubsectionfont{\bf}
\setstretch{1.125} 
\setlength{\skip\footins}{0.8cm}
\setlength{\footnotesep}{0.25cm}
\setlength{\jot}{10pt}
\titlespacing*{\section}{0pt}{4pt}{4pt}
\titlespacing*{\subsection}{0pt}{15pt}{1pt}

\fancyfoot{}
\fancyfoot[LO,RE]{\vspace{-7.1pt}\includegraphics[height=9pt]{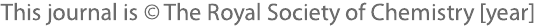}}
\fancyfoot[CO]{\vspace{-7.1pt}\hspace{13.2cm}\includegraphics{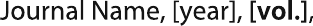}}
\fancyfoot[CE]{\vspace{-7.2pt}\hspace{-14.2cm}\includegraphics{head_foot/RF}}
\fancyfoot[RO]{\footnotesize{\sffamily{1--\pageref{LastPage} ~\textbar  \hspace{2pt}\thepage}}}
\fancyfoot[LE]{\footnotesize{\sffamily{\thepage~\textbar\hspace{3.45cm} 1--\pageref{LastPage}}}}
\fancyhead{}
\renewcommand{\headrulewidth}{0pt} 
\renewcommand{\footrulewidth}{0pt}
\setlength{\arrayrulewidth}{1pt}
\setlength{\columnsep}{6.5mm}
\setlength\bibsep{1pt}

\makeatletter 
\newlength{\figrulesep} 
\setlength{\figrulesep}{0.5\textfloatsep} 

\newcommand{\topfigrule}{\vspace*{-1pt}%
\noindent{\color{cream}\rule[-\figrulesep]{\columnwidth}{1.5pt}} }

\newcommand{\botfigrule}{\vspace*{-2pt}%
\noindent{\color{cream}\rule[\figrulesep]{\columnwidth}{1.5pt}} }

\newcommand{\dblfigrule}{\vspace*{-1pt}%
\noindent{\color{cream}\rule[-\figrulesep]{\textwidth}{1.5pt}} }

\makeatother

\twocolumn[
  \begin{@twocolumnfalse}
\vspace{3cm}
\sffamily
\begin{tabular}{m{4.5cm} p{13.5cm} }
\includegraphics{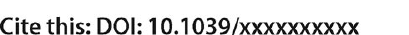} & \noindent\LARGE{\textbf{Propagating Density Spikes in Light-Powered Motility-Ratchets
}}\\ 
\vspace{0.3cm} & \vspace{0.3cm} \\

& \noindent
Celia Lozano \textit{$^{a,\ast}$}, 
Benno Liebchen \textit{$^{b,c,\dagger,\ast}$}, 
Borge ten Hagen \textit{$^{b,d,\ast}$}, 
Clemens Bechinger \textit{$^{a}$} and 
Hartmut L\"{o}wen\textit{$^{b}$} 
\\

\includegraphics{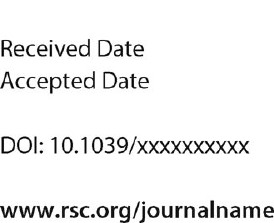} 

& \noindent\normalsize{
Combining experiments and computer simulations, we use a spatially periodic and flashing light-field to
direct the motion of phototactic active colloids. 
Here, the colloids self-organize into 
a density spike pattern, which resembles a shock wave and propagates over long distances, almost without dispersing. 
The underlying mechanism involves a 
synchronization of the colloids with the light-field, so that particles see the same intensity gradient each time the light-pattern 
is switched on, but no gradient in between (for example).  
This creates a pulsating transport whose strength and direction can be controlled via the flashing protocol and the self-propulsion speed of the colloids. 
Our results might be useful for drug delivery applications and can be used to segregate active colloids by their speed. 
}
\end{tabular}
\end{@twocolumnfalse} \vspace{0.6cm}
]

\renewcommand*\rmdefault{bch}\normalfont\upshape
\rmfamily
\section*{}
\vspace{-1cm}

\footnotetext{\textit{$^{a}$~Fachbereich Physik, Universität Konstanz, Konstanz  78457, Germany.}}
\footnotetext{\textit{$^{b}$~Institut f\"{u}r Theoretische Physik II: Weiche Materie, Heinrich-Heine-Universit\"{a}t D\"{u}sseldorf, 40225 D\"{u}sseldorf, Germany.}}
\footnotetext{\textit{$^{c}$~Institut f\"ur Festk\"orperphysik, Technische Universit\"at Darmstadt, 64289 Darmstadt, Germany.}}
\footnotetext{\textit{$^{d}$~Physics of Fluids Group and Max Planck Center Twente, Department of Science and Technology, MESA+ Institute, and J. M. Burgers Centre for Fluid Dynamics, University of Twente, 7500 AE Enschede, The Netherlands.}}
\footnotetext{\textit{$^{\dagger}$~corresponding author: liebchen@fkp.tu-darmstadt.de}}
\footnotetext{\textit{$^{\ast}$~These authors contributed equally.}}

\section{Introduction}
\label{sec:intro}
Active colloids are autonomously navigating microparticles that consume energy while moving. They comprise 
living microorganisms like bacteria, algae and sperm \cite{Romanczuk2012,Elgeti2015,Cates2012}, but also 
man-made synthetic swimmers, which can be produced with desired properties. 
Such synthetic microswimmers are often based on anisotropic colloidal Janus particles that are self-propelled by phoretic mechanisms 
either directly induced by catalytic surfaces \bn{evoking chemical reactions} \cite{Paxton2004,Palacci2010}, or initiated by light 
\cite{Volpe2011,Buttinoni2012,Palacci2013JACS,Palacci2013Science,Palacci2014,Moyses2016,Schmidt2018} or other external fields, 
such as ultrasonic \cite{Wang2012}, 
magnetic \cite{Dreyfus2005,Grosjean2015,Steinbach2016,Kaiser2017} or electric \cite{Bricard2013,Morin2017} ones. 

While free active colloids show a diffusive random motion on large scales \cite{Howse2007,Bechinger2016}, 
equivalent to the motion of passive colloids at high temperature, 
applications to use them e.g. for targeted drug delivery \cite{Gao2012,Park2017} or nanorobotics \cite{Hong2010}
require to direct and steer their motion on demand. 

One way to direct the motion of active particles is to expose them to a periodic but asymmetric potential landscape (ratchet), leading to directed transport 
\cite{Angelani2011,Pototsky2013,Reichhardt2017,Zampetaki2018}, 
in a way similar as for passive colloids driven out of equilibrium through additional time-dependent fields \cite{Reimann2002,Hanggi2009,Mukhopadhyay2018}.
Characteristically, such potential ratchets involve forces acting on the center of mass coordinate of the particles, yielding a spatial variation of their potential energy. 
A versatile alternative to create directed transport in active colloids, are so-called motility-ratchets, which specifically exploit the active nature of the particles and have no 
direct counterpart for passive colloids. These ratchets 
hinge on the spatial modulation of the self-propulsion speed (or direction) 
through an external field \cite{Lozano2016,Stenhammar2016,Reichhardt2017}, not affecting the potential energy of the particles. 
Here, the required modulation of the self-propulsion speed 
can be achieved e.g. for 
light-sensitive Janus colloids in a suitable standing light-wave 
\cite{Lozano2016} which has been previously discussed in the context of dynamical trapping of active particles in the dark spots of the light field \cite{Magiera2015,Grauer2018}.
An interesting extension of such static motility-ratchets, providing an additional handle to control the active particle dynamics, 
is to use a time-dependent motility field, 
as recently considered 
theoretically \cite{Geiseler2016PRE,Geiseler2017Entropy,Geiseler2017SciRep}, and also experimentally \cite{Koumakis2018} for 
light-sensitive bacteria \cite{Arlt2018,Frangipane2018,Arlt2019}.

\begin{figure}
\centering
   \includegraphics[width=0.48\textwidth]{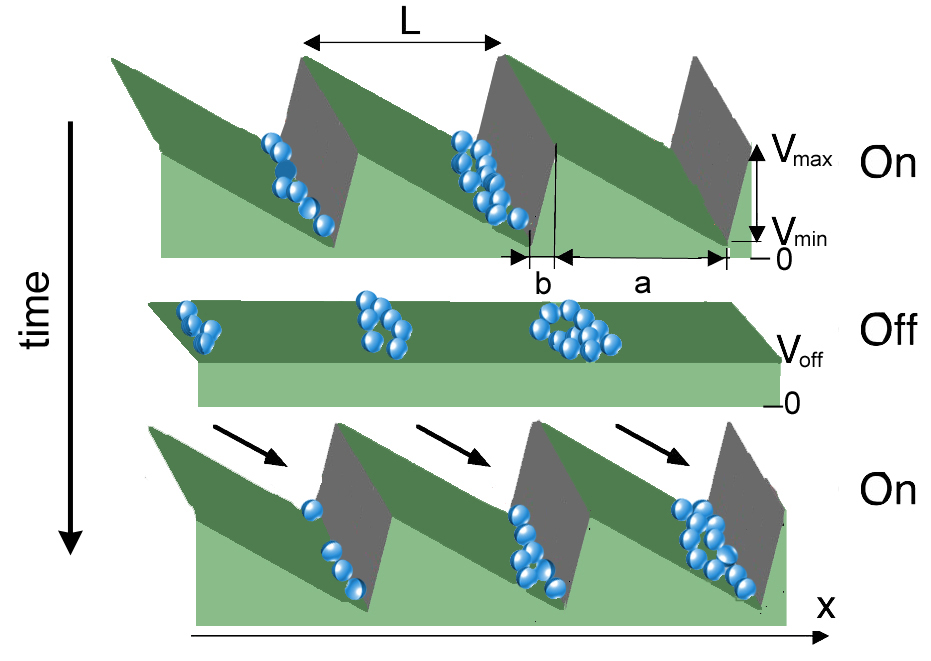} 
  \caption{Schematic: Active colloids polarize and self-organize into density spikes propagating resonantly through the flashing light field. 
Here, the colloids synchronize with the flashing light-field and a macroscopic fraction of them essentially follows the same periodic trajectory, as dictated by a limit cycle of the system,
so that particles within each density spike move coherently. 
This creates a pulsating particle transport with low dispersion. 
The shown sawtooth-shaped profile represents the particle's self-propulsion velocity, 
which varies between $v_{\rm min}$ and $v_{\rm max}$ when the light-pattern is on (``on-phase''), where 
$a,b$ determine the steepness of the gradient, 
and which everywhere equals $v_{\rm off}$, when the light-field is uniform (middle panel, ``off-phase'').}
  \label{fig:profile}
\end{figure}

In the present work, we combine simulations and experiments to establish a flashing motility-ratchet for synthetic phototactic colloids, 
based on a sawtooth-shaped light pattern with intensity $I(x,t)$ which we periodically switch on and off (flashing). 
The gradients of $I$ create an effective torque affecting the colloids' self-propulsion direction which can systematically bias their motion \cite{Bickel2014,Lozano2016}, 
yielding a directed transport. Here, the emergence of transport hinges on a phototactic torque biasing the active particles' direction of motion, as opposed to classical flashing potential ratchets
\cite{Reimann1997,Hanggi2009}, usually depending on the interplay of time-dependent forces and diffusion.
This novelty in the working mechanism of the flashing motility-ratchet
manifests in a set of remarkable features. In particular, we find that 
the individual active particles synchronize with the flashing light-field and self-organize into 
density-spikes resembling a coherently moving shock-wave. As their most striking feature, these spikes hardly disperse, 
opposing the usual situation in potential ratchets, where the 
interplay of time-dependent forces and diffusion
leads to strong dispersion of any localized particle ensemble. 
Thus, the present setup opens a route to use laser-light to create a pulsating transport allowing to 'bombard' a distant target with short and intense pulses of active particles, 
as might be interesting, in particular, for 
drug-delivery applications.  
Here, the transport velocity can be systematically controlled via the flashing times, but interestingly, it
also depends sensitively on the particle speed in the off-phase. 
In particular, we find that the 
transport direction even reverts when changing the 
self-propulsion speed of the particles in the off-phase. This transport reversal can be used, in principle, to  
segregate ensembles of fast and slow particles and might serve as a useful tool for the preparation of clean
ensembles \bn{of} active particle with near identical self-propulsion speed. 

Before detailing these findings, let us sketch the physical mechanism underlying the flashing motility-ratchet: 
If the torque acting on the active colloids scales linearly with the gradient of the laser field (unsaturated regime) no transport can occur in 
static light patterns. (This contrasts \cite{Lozano2016} operating in the saturated regime.) Intuitively, if noise is negligible, this is because the phototactic torque acting on 
particles crossing a whole spatial period, first turns them into a certain direction and then back to the original orientation, 
so that a localized and unbiased initial ensemble remains unbiased for all times. 
Flashing in turn allows the particles to synchronize with the light field, in a way that they repeatedly see the same gradient 
when the light pattern is on and a uniform field in each off-phase, provoking a persistent unidirectional motion. This dynamics
is based on a limit cycle attractor in the underlying phase space, which 
represents a late-time dynamics where particles 
move by exactly one spatial period per flashing cycle, in suitable parameter regimes. 
Since all particles 
which are attracted by the same limit cycle show 
one and the same periodic dynamics at late times they move coherently with a speed dictated by the limit-cycle, 
leading to an almost dispersion-free transport - a key feature of the present work. 

\section{Model}\label{sec:model}
For conceptual clarity, we first introduce an idealized flashing motility ratchet, 
based on an effective phototactic torque which scales linearly with the 
light gradient ($\omega \propto |\nabla I|$). In this case, the emerging transport is flashing-induced and 
vanishes in static light patterns.
To see this, consider self-propelled Janus particles, actively moving in 2D with a self-propulsion speed $v(x,t)$, varying both in space and time, 
as controlled by the imposed light-field (see Fig.~\ref{fig:profile}). For simplicity, we specifically
consider a quasi 1D modulation of the light field, and hence of $v$. 
The self-propulsion direction $\hat{\mathbf{u}}=(\cos\phi,\sin\phi)$
changes in response to an effective phototactic torque, and also due to rotational diffusion, yielding:
\begin{eqnarray}
\dot{\mathbf{r}} & = v(x,t) \hat{\mathbf{u}} + \sqrt{2D_t}\boldsymbol{\zeta}_{\mathbf{r}}(t), \label{tdyn} \\
\dot{\phi} & = \omega(x,\phi,t) + \sqrt{2D_r}\zeta_{\phi}\bn{(t).} \label{odyn}
\end{eqnarray}
Here ${\bf r}=(x,y)$ and $D_t,D_r$ are translational and rotational diffusion coefficients; $\boldsymbol{\zeta}_{\mathbf{r}}(t)$ and $\zeta_{\phi}(t)$ represent
Gaussian white noise of zero mean and unit variance. 
The key-quantity controlling the particle dynamics in the light-field is the phototactic alignment rate $\omega$, which reads 
\begin{equation}
\omega(x,\phi,t)= 
A v(x,t) v'(x,t) \sin\phi
\label{unsaturated}
\end{equation}
where $v'(x,t)=\partial v(x,t)/\partial x$.
Eq. (\ref{unsaturated}) represents a linear relationship between alignment rate and intensity gradient, $\omega \propto |\nabla I|$, which is realistic for 
shallow light patterns \cite{Lozano2016}, but will later be generalized towards saturation effects. 
Here, the coefficient $A$ follows from experiments \cite{Lozano2016}. 
For the velocity profile $v(x,t)$, we choose a sawtooth-shape in the on-phase, as sketched in Fig.~\ref{fig:profile}, with segment \bn{lengths} $a,b$
and minimal and maximal velocities of 
$v_{\rm min}$ and $v_{\rm max}$ respectively. In the off-phase the velocity is uniform $v(x,t)=v_{\rm off}$.
\\Note that in general, besides creating an effective torque aligning the particles, light gradients 
also induce effective forces creating particle translations, represented by a term $\propto \nabla I$ on the r.h.s of Eq.~(\ref{tdyn}). 
In accordance with \cite{Lozano2016}, we here neglect such a term for simplicity, but emphasize its existence for future reference.

\section{Flashing-induced Coherent Transport}\label{sec:results}
\begin{figure}
\centering
	\includegraphics[width=0.49\textwidth]{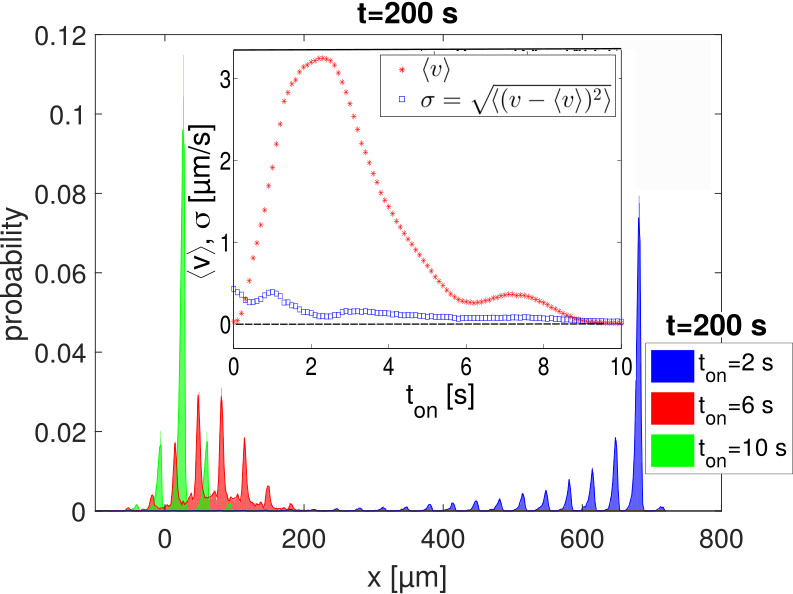}
  \caption{Propagating density spikes: Spatial particle distribution after $t=200s$ for $N_p=50.000$ and different $t_\text{on}$ shown in the key. 
Inset: Mean transport velocity $\langle v \rangle$ and standard deviation $\sigma = \sqrt{\langle(v-\langle v \rangle)^2\rangle}$ 
as a function of $t_\text{on}$, calculated for $N_p=100$ and $t_{\rm end}=30T$ per data point.
The dashed line shows the zero line to guide the eye.
Parameters: $t_\text{on}+t_\text{off}=10s$; $a=27.38\mu m; b=6.02 \mu m; v_{\rm min}=3 \mu m/s; v_{\rm max}=13 \mu m/s$ and $v_{\rm off}=3\mu m/s$; $A \approx 0.65$ 
($A=C_1 C_2/R$ for particles with radius $R=1.365\mu m$ and $C_1=0.44$, $C_2=2.01s$ being experimentally determined parameters, see \cite{Lozano2016}).}
  \label{fig:t_on_unsat}
\end{figure}

Let us now explore the dynamics of a representatative particle ensemble in the flashing light-field (Fig.~\ref{fig:profile}). 
We choose random initial positions and orientations uniformly distributed within one 
unit cell of the sawtooth-shaped light pattern ($x\in [0,L)$ and $\phi \in [0,2\pi)$) and
define the average transport velocity as $\langle v\rangle=\lim\limits_{t \rightarrow t_{\rm end}}{[x(t)-x(0)]/t}$, 
where $t_{\rm end}$ is some time, large enough that $\langle v\rangle$ is basically stationary.  
Now performing Brownian dynamics simulations (see \footnote{Brownian dynamics simulations have been performed using a standard forward Euler algorithm with a time step of $dt=0.001T$. Each point in curves showing averages is based on $N_p=100$ trajectories (initial conditions), integrated for $30T$ and distributions are based on $N_p=50.000$ initial conditions. The system size is infinite, i.e. the simulations do not require any boundary conditions.} for details) until $t=t_{\rm end}$ for various $t_{\rm on}$ but fixed 
flashing period $T=t_{\rm on}+t_{\rm off}$, we generically find a directed transport for all values of $t_{\rm on}$ (Fig.~\ref{fig:t_on_unsat},
inset). In contrast, however, for the static cases $t_{\rm on}=0$ and $t_{\rm off}=T$ (uniform light field) and 
$t_{\rm on}=T$ and $t_{\rm off}=0$ (static sawtooth-shaped light pattern) the transport vanishes.
That is the directed transport is flashing-induced. 

Before exploring the origin and properties of the flashing-induced transport, let us first understand its absence in uniform light fields. 
To see this, note that particles crossing a whole spatial period of a static light pattern ($t_{\rm on}=10s$), 
do not experience any net alignment, which can be seen as follows for the
noise-free case: 
from Eqs.~(\ref{tdyn},\ref{odyn},\ref{unsaturated}) and $\dot \phi=\dot x (\partial \phi/\partial x)$, 
it follows that for vanishing noise we have 
$\partial \phi/\partial x=A v'(x) \tan\phi(x)$, which yields, after integration over 
a spatial period $\phi(x+L)=\phi(x)$. Thus, in the absence of noise, if all particles are initialized with uniformely distributed orientations 
in an intensity-maximum of the light field, exactly half of them \emph{permanently} move to the left and to the right respectively, 
i.e. there is no transport.
(For spatially distributed initial ensembles, we provide a slightly more general 
argument for the absence of transport in footnote \cite{footnote1}.)
Our argument for the absence of transport breaks down in the presence of flashing.  
Intuitively, flashing allows the particles to synchronize with the light field in a way that they 
``see'' the same gradient each time the light-field is switched on and no 
gradient when the light-field is off. This repeatedly aligns their self-propulsion in the same direction, as we further detail in section 4. 

The synchronization of the particle motion with the light-field has a number of remarkable consequencies. One of them 
is shown in Fig.~\ref{fig:t_on_unsat} (inset), where we can see that the standard deviation
$\sigma=\sqrt{\langle v^2\rangle-\langle v\rangle ^2} \ll \langle v \rangle$, for most values of $t_{\rm on}$. 
That is, the transported particle ensemble hardly disperses, but moves rather coherently through the flashing light field. 
In particular, $\sigma$ has a minimum at $t_{\rm on} \approx 2.6s$ where the 
transport velocity is maximal.  
To further illuminate the particle dynamics in the flashing motility-ratchet, 
let us consider the spatial particle distribution in the light-pattern after, a snapshot of which after 20 driving cycles is shown in Fig.~\ref{fig:t_on_unsat}. 
As we can see in this figure, for $t_{\rm on}=10s$, where we have no transport, 
most particles are localized around their initial positions; they collectively move back and forth in the light-pattern. 
Conversely, for $t_{\rm on}=2s$ the particle distribution is strongly asymmetric and shows a train of pronounced peaks 
resembling a shock wave. This train persistently moves to the right, with a speed of precisely one spatial periode per flashign cycle ($v=L/T$).

\begin{figure}
\centering
  \includegraphics[width=0.4\textwidth]{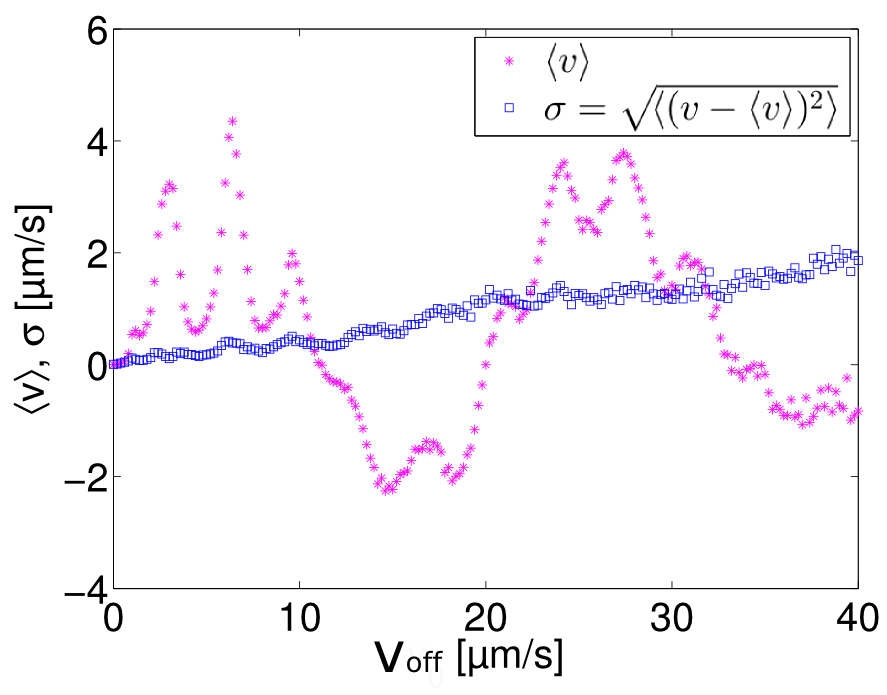} 
\caption{Transport reversal: Mean velocity $\langle v \rangle$ and standard deviation $\sigma = \sqrt{\langle(v-\langle v \rangle)^2\rangle}$ for $t_{\rm on}=2s$ as a function of $v_{\rm off}$. 
Other parameters as in Fig.~\ref{fig:t_on_unsat}.}
  \label{fig:v0dep}
\end{figure}

Apart from exploring the dependence of the transport velocity on parameters of the environment, such as the flashing duration $t_{\rm on}$, it 
is instructive to also explore the dependence of the transport velocity on particle properties: 
In Fig.~\ref{fig:v0dep}, we show the transport velocity as a function of $v_{\rm off}$ (self-propulsion speed in the off state), finding a remarkable structure, comprising distinct peaks in the 
transport velocity, and in particular a transport-reversal for $10\mu m/s \lesssim v_{\rm off} \lesssim 20\mu m/s$. (Note here, that the structure of the peaks is statistically well converged, as indicated by the fact that the standard deviation $\sigma$ is much smaller than $\langle v\rangle$ for most data points.)
Following the observed transport reversal, when considering a mixture of particles with different velocities, say $5\mu m/s$ and $15 \mu m/s$, the flashing light-field 
will transport them in opposite directions and segregate the mixture. 
This could be used in principle 
as a tool to prepare clean ensembles of active particles with a uniform 
self-propulsion speed. 
Notice that 
the presence of a current reversal alone, which is naturally present in many potential ratchets \cite{Reimann2002}, 
is not sufficient to achieve a clean and near-complete segregation of two initially distributed particle species. Instead, such a 
segregation requires $|\langle v \rangle| \gtrsim \sigma$ for each species, a condition which is naturally fulfilled for the 
flashing motility-ratchet (Fig.~\ref{fig:v0dep}). This shows once more that 
the intrinsically low dispersion of the emerging transport might be useful for practical applications.
Notice however, that  
$\sigma$ increases roughly linearly with $v_{\rm off}$ (Fig.~\ref{fig:v0dep})
i.e. with the distance particles travel within each \bn{off-phase}. That is, the particle distribution gets broader as $v_{\rm off}$
increases allowing for a clean segregation 'only' for particles which are not too fast.

\section{Mechanism}
\begin{figure}
\centering
  \includegraphics[width=0.5\textwidth]{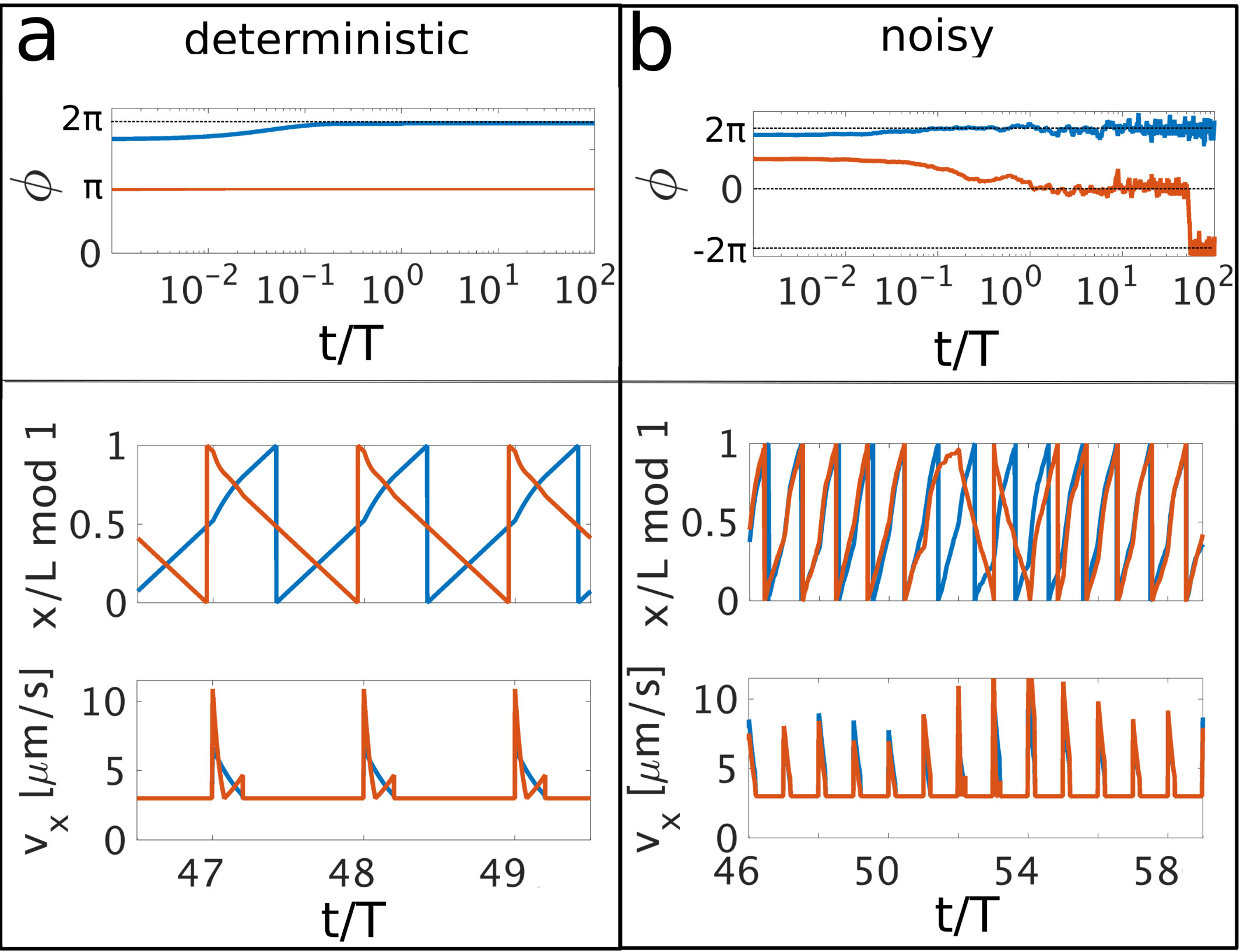} 
\caption{Transport mechanism in the flashing motility ratchet without (a) and with (b) noise for $t_{\rm on}=2s$ and two representative trajectories (blue and orange) with initial orientations $\phi=3.1$ and $\phi=5.5$. 
Without noise, particles fully align with the symmetry axis of the light-field (${\bf p} \times \nabla I=0$), yielding 
$\phi=\pi+2\pi n$ or $\phi=2\pi n$ ($n\in \mathbb{Z}$), representing 
motion to the left and to the right, respectively.
At late times (middle and lower panel), particles reach a limit cycle attractor and move periodically, with an average speed of exactly $v=L/T$ (middle panel) either to the left or to the right. 
In the presence of noise, the particle orientations do not fully converge but fluctuate. Here 
particles show a dynamics which is similar to the noise-free dynamics most of the time (middle panel), but they can cross-over from one attractor to another one, which can, for example, lead to temporary 
motion into the opposite direction (middle panel). The shown time intervals have been chosen to 
representatively illustrate the described dynamics (periodic motion and attractor hopping).
Parameters as in Fig.~\ref{fig:t_on_unsat}.}
  \label{fig:mech}
\end{figure}

To understand the mechanism leading to a coherently travelling density spike pattern in detail, it 
is instructive to first explore typical trajectories in the noise-free case ($D_r=D_t=0$), as visualized in 
Fig.~\ref{fig:mech}a for the case $t_{\rm on}=2s$.
Here, within one or several driving cycles, the orientation of the particles, which align antiparallel to the light-gradient, converges to a steady value of either $\phi=0$ (or $\phi=2\pi n$, $n$ being integer) or to $\phi=(2n+1)\pi$
(particles typically need about $2-4$s to fully turn \cite{Lozano2016}) 
corresponding to perfect 
alignment with the axis along which the light-pattern is modulated 
(upper panel). 
Once they are aligned, particles no longer experience a torque from the light-field (${\bf p} \times \nabla I(x,t)=0$ for all $x,t$) and persistently self-propel in one and the same direction.
Conversely to their constant orientation, the speed of the particles still evolves rather irregularly at this stage: it simply reads $v=v_{\rm off}$ in each \bn{off-phase}, 
but in the \bn{on-phase} $v \in (v_{\rm min},v_{\rm max})$ it 
depends on the particle's position at the instance the light-field is 
switched on. 
At late times, the dynamics 
converges to a limit cycle attractor, whose existence is expected from the Poincar\'e-Bendixon theorem, resulting in a regular, periodic dynamics (Fig.~\ref{fig:mech}a, middle and 
lower panel).
That is particles fully synchronize with the flashing light-field with a temporally periodic speed, whose average 
is exactly $v=L/T$ (middle panel), i.e. particles propagate by exactly one spatial period $L$ per $T=10s$.
As illustrated by the blue and the orange curve in Fig.~\ref{fig:mech}a (middle panel), depending on the initial conditions, particles follow a corresponding orbit either to the left or to the right. 
In this minimal case where only two limit cycles are relevant, each particle propagates the same one-flashing-cycle-averaged velocity of $v=L/T$ either to the left or to the right (Fig.~\ref{fig:mech}a, middle and 
lower panel). 
The resulting late-time transport velocity is then fully determined by the ratio of initial conditions reaching the limit cycle corresponding to motion to the 
left and to the right, respectively, i.e. to the 'basins of attraction' of the two limit cycles. Specifically for the considered case of $t_{\rm on}=2s$, 
most of the \bn{particles} reach an attractor pointing to the right, which generates the transport.
(In general, depending on $t_{\rm on}$ besides limit cycles representing a 1:1 resonance, also limit cycles
allowing particles to traverse $n$ spatial periods within $m$ driving cycles are allowed, leading \bn{to} a particle speed of $v=nL/(m T)$.)

In the presence of noise, 
$\phi$ does of course not fully converge to $\pi$ or $2\pi$ but fluctuates even at late times (upper panel in Fig. \ref{fig:mech}b). 
However, most of the time (for $t_{\rm on}=2s$), particles still follow a near-resonant dynamics and move with $v\approx L/T$ 
(middle panel in Fig. \ref{fig:mech}b), temporarily resembling trajectories of the underlying deterministic system. Such a dynamics can
prolong for many driving cycles. 
From time to time, however, noise relocates a particle from one limit cycle of the underlying deterministic system to another one, leading, at least temporarily to a different dynamics. 
The dynamics shown in Fig. \ref{fig:mech}b (middle panel) illustrates this: In the time interval \bn{between} $t\sim 510s$ and $t\sim 550s$, the 
orange trajectory crosses from the ``$\phi=0$-attractor'' of the underlying noise-free system over to 
the ``$\phi \approx -2\pi$ attractor'', yielding the 
same dynamics, but in between, it follows the ``$\phi=-\pi$-attractor'' for a few flashing cycles and moves into the opposite direction.  
In the presence of noise, the resulting transport velocity is therefore mainly determined by the time particles spend in the basin of attraction of the limit cycles of the underlying noise-free system. When  
this basin of attraction is small, as e.g. for $t_{\rm on}=2s$ for the ``$\phi =(2n+1)\pi$-attractor'' which leads to motion to the left, particles all move to the right in the long-time average.  

The spike pattern observed above now follows quite naturally from the described particle-light-field synchronization. 
Here, each spike represents a package of particles which have followed the same limit cycle dynamics for the same amount of time. In particular, 
particles which have moved for $200s=20T$ with a velocity of $v\approx L/T=33.6\mu m/10s$, have traversed a distance of $x\approx 20L \approx 672\mu m$, which corresponds to the 
first large peak in Fig.~\ref{fig:t_on_unsat} for $t_{\rm on}=2s$; the following peaks \bn{correspond} to particles having traversed a distance of $x\approx 19L,18L,17L..$, corresponding to particles which have either initially
'lost' time by aligning with a certain delay, or by moving temporarily into the opposite direction. 
Finally, there is a small peak of particles having traversed a distance $x>20L$; this corresponds to particles which have been pushed forward 
during the on-phases before synchronizing with the light-field.  
(The spike pattern is less clean for $t_{\rm on}=6s$ and $t_{\rm on}=10s$, where the relevant attractors do not correspond to 1:1 resonances.)

\section{Torque saturation}
\begin{figure}
\begin{center}
\includegraphics[width=0.5\textwidth]{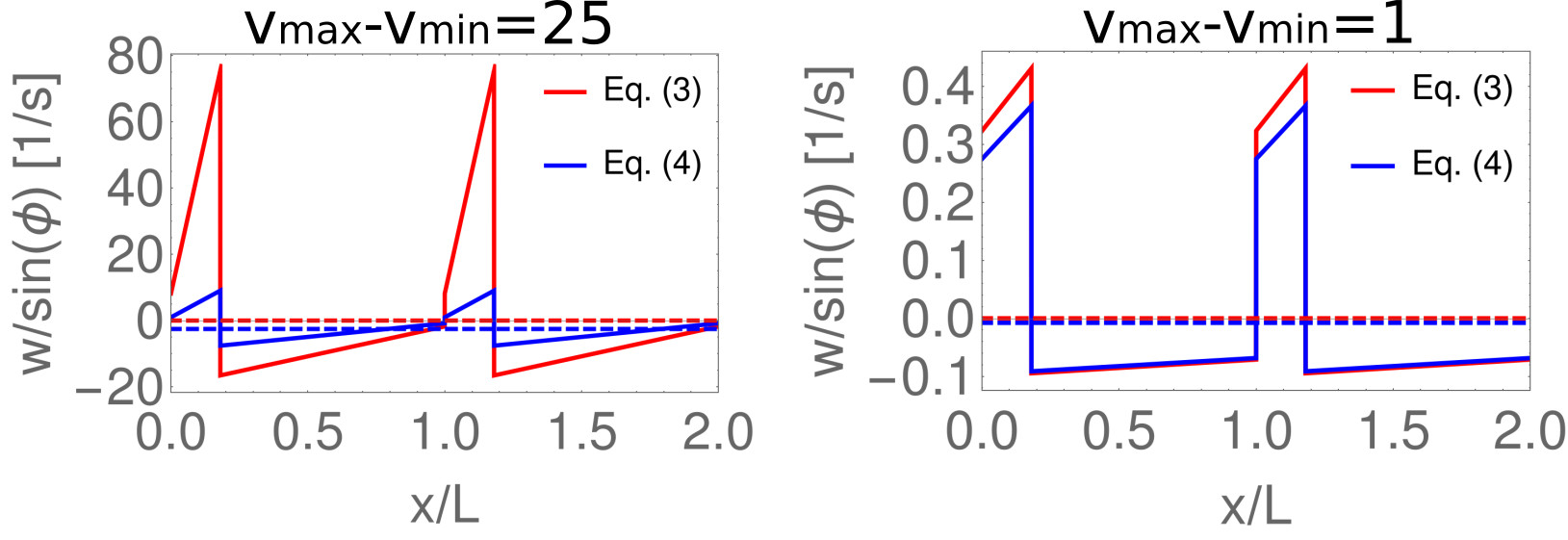}
\caption{Red and blue curves represent Eqs. (\ref{unsaturated}) and (\ref{omegasat}), respectively, deep in the saturated (left) and in the unsaturated (right) regime. Dashed lines show averages of full curves. Parameters as in Fig.~\ref{fig:t_on_unsat}}
\label{fig_eqs34}
\end{center}
\end{figure}

\begin{figure}
\centering
  \includegraphics[width=0.4\textwidth]{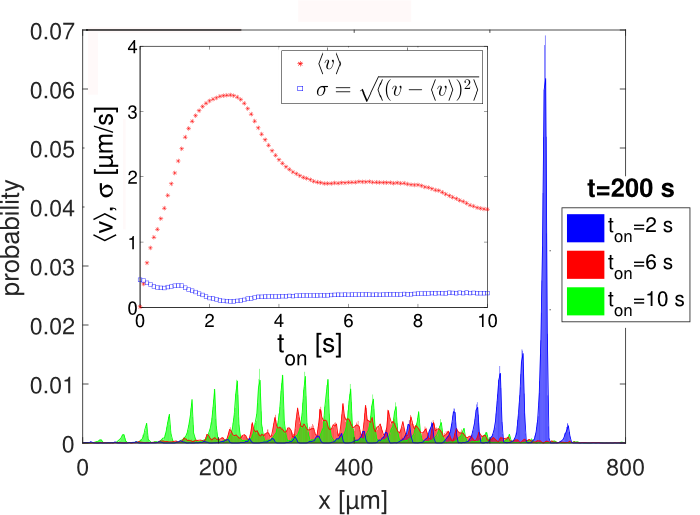} \\
  \caption{Impact of torque saturation on particle distribution and transport velocity. 
Figures and parameters as Fig.~\ref{fig:t_on_unsat} but in the presence of torque saturation, Eq.~(\ref{omegasat}),
($C_1=0.44; R=1.365\mu m; C_2=2.01s$, obtained from experiments).}
  \label{fig:t_on_sat}
\end{figure}

In experiments the phototactic alignment rate of a Janus colloid with the intensity gradient does not generally increase linearly, 
as we have assumed so far, but saturates for steep intensity gradients due to thermal coupling \cite{Lozano2016}. 
Thus, to compare the flashing motility ratchet with experiments in the next section, we wish to understand the impact of saturation effects first. 
To this end, we use the following expression for the 
phototactic alignment rate \cite{Lozano2016}
\begin{equation}
\omega(x,\phi,t)=v(x,t) \sin\phi \sgn\left(v'(x,t)\right) \frac{C_1}{R} \left(1-\exp\left(-C_2 \left|v'(x,t)\right|\right)\right)
\label{omegasat}
\end{equation}
which approximately reduces to Eq.~(\ref{unsaturated}) for $\left|v'(x,t)\right| \ll 1/C_2$ and $A=C_1 C_2/R$ (and exactly for $|v'(x,t)|\rightarrow 0$). See Fig.~\ref{fig_eqs34} for a comparison of Eqs.~(\ref{unsaturated}) and (\ref{omegasat}), both in the saturated and in the unsaturated regime.
Here, $C_2$ controls the crossover from the linear to the saturated region. 
To see the impact of saturation effects, let us now explore the transport velocity as a 
function of $t_{\rm on}$ (\bn{Fig.}~\ref{fig:t_on_sat}, inset) for $C_2=2.01s$ as previously determined in \cite{Lozano2016}. Just as for the unsaturated case (Fig.~\ref{fig:t_on_unsat}), 
the transport increases about linearly for small $t_{\rm on}$, reaches a maximum at $t_{\rm on} \approx 2.6s$ of 
$\langle v\rangle\gtrsim 3 \mu m/s$ and then decays again, suggesting that saturation effects 
have little bearing on the flashing motility ratchet. 
However, at larger values of $t_{\rm on}$, the transport plateaus (Fig.~\ref{fig:t_on_sat}), rather
than decaying towards zero and reaches a finite value of $\langle v\rangle\gtrsim 1.5 \mu m/s$ for $t_{\rm on}\rightarrow 10s$. That is, saturation effects create a directed transport even 
for a stationary (non-flashing) light field. 
This special case has been previously explored in \cite{Lozano2016}. 
Why do we obtain a directed particle transport here even in the absence of flashing? Note first, that the argument for the 
absence of transport in a static light pattern as given in 
section 3 breaks down in the presence of torque saturation, as $\omega(x,\phi)$ is no longer spatially periodic and a particle 
crossing a whole spatial periode can therefore experience a net change of its orientation (even in the absence of noise). 

More constructively, on average, particles need significantly more time to cross the 
long and shallow $b$-segments where $\nabla I$ is small and the alignment rate is essentially unsaturated (i.e. $\omega \propto |\nabla I|$) 
than for crossing the short and steep $a$-segments, where $\omega$ saturates and is only slightly larger than in the $b$-segments.
Hence, particles leaving a steep $a$-segment are aligned only weakly as compared to those leaving a $b$-segment.
Thus, particles leaving the $a$-segement to the left are commonly reflected in the adjacent $b$-segment, since 
phototaxis opposes their swimming direction. Conversely, particles leaving 
an $a$-segment to the right almost certainly manage to cross the adjacent $b$-segement where phototaxis is supportive and speeds them up. 
This breaks the left-right symmetry, initiating a transport to the right. 
(In the extreme case of an almost vertical $a$-segement and an extremely flat and long $b$-segment, basically all particles would get reflected, 
when leaving an $a$-segement to the left, and the ratchet would serve as an ``active particle diode''.)

To further illuminate the impact of saturation effects, let us 
explore the particle distribution in the lattice, say after 20 driving cycles. From Fig.~\ref{fig:t_on_sat}, we can see that the distribution for $t_{\rm on}=2s$
is essentially the same as in the absence of saturation effects (Fig.~\ref{fig:t_on_unsat}), preserving the shock-wave like profile. 
As for the transport itself, however, the distribution significantly deviates from the unsaturated case when $t_{\rm on}\sim T$.
Following these observations, for our upcoming experiments we expect that saturation effects do not play a strong role if 
$t_{\rm on} \ll T$. However, if 
$t_{\rm on} \sim T$, saturation effects are expected to significantly impact the transport.

\begin{figure*}
\centering
  \includegraphics[width=0.98\textwidth]{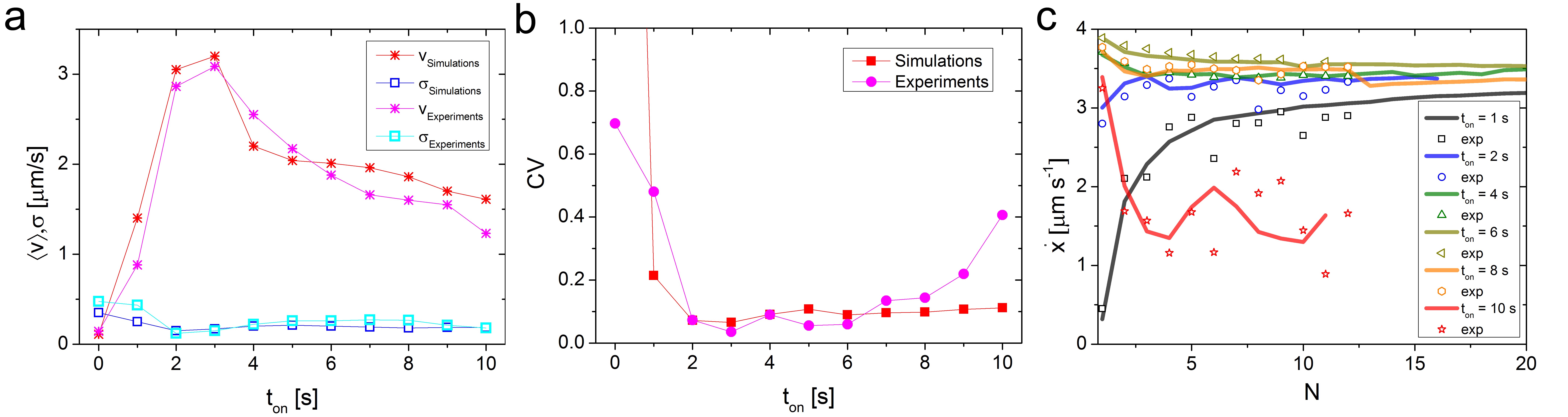} \\
  \caption{Comparison of model and experiments: (a) Average velocity $\langle v \rangle$ and standard deviation $\sigma$, (b) coefficient of variation ($CV=\langle v\rangle /\sigma$), averaged over the entire simulation/experiment (see text), as a function of 
$t_{\rm on}$. Panel (c) shows the average particle velocity in the $N$-th flashing cycle $\langle \dot x \rangle$ for different $t_{\rm on}$ shown in the key.
Parameters, both in experiments and simulations $T_\text{f}=t_{\rm on}+t_{\rm off}=10s$, 
$v_{\rm max}=8\mu m/s, v_{\rm off}=4\mu m/s; v_{\rm min}=0.8\mu m/s$ and 
$C_1=0.44; R=1.365\mu m; C_2=2.01s$ from fits to experiments in \cite{Lozano2016}.
}
  \label{fig:exp}
\end{figure*}
\section{Experiments}
To test the flashing motility-ratchet, we now compare our results with experiments. 
We use light-activated Janus colloids, which are composed of optically transparent silica spheres ($R=1.36\mu m$) being 
capped on one side with $20nm$ of carbon. The active colloids are suspended in a critical mixture of water and 2,6-lutidine 
(lutidine mass fraction 0.286), whose lower critical point is at a temperature of $T_c = 34.1^\circ C$. 
When the solution is kept at a temperature of $30^\circ C<T_c$, the colloids perform diffusive Brownian motion. 
Upon laser illumination (at wavelength $\lambda =532nm$), where light is absorbed at the caps only, the solvent locally demixes. This  
creates a concentration gradient accross each colloid's surface, leading to self-propulsion. The resulting self-propulsion speed 
scales linearly with the laser intensity \cite{Gomez2017}.
The light pattern, and hence also the motility-pattern, is created by a laser line focus ($\lambda=532nm$) 
being scanned across our sample cell by means of a galvanostatically driven mirror 
with a frequency of $200Hz$.
Synchronization of the scanning motion 
with the input voltage of an electro-optical modulator leads to a quasi-static illumination landscape. 
The alternation between an instensity ladscape and an homogenous one were fully automated using a 
customized software written in LABVIEW. Since the remixing timescale of the binary mixture 
is on the order of $100ms$ \cite{Gomez2017}, 
the periodic mirror motion is fast enough to produce stable particle self-propulsion. Particle positions and orientations were obtained by digital video microscopy with a frame rate of 13 fps.

We use a very dilute suspension of microswimmers to avoid 
particle interactions and subject it to the described flashing ratchet. 
Here we track trajectories of particles initialized 
in one and the same $a$-segment with uncontrolled initial orientations and let them evolve for 13$T$. We repeat the experiment several times and average over $N_p>25$ trajectories. 
In Fig.~\ref{fig:exp}a we compare the resulting transport velocity in experiments with our model for various flashing times, 
finding close quantitative agreement.
For the considered parameters, the transport velocity rapidly increases from $\langle v\rangle=0$ at $t_{\rm on}=0s$
and approaches a maximal speed of about $\langle v\rangle \gtrsim 3\mu m/s$ for $t_{\rm on}\sim 2-3s$, 
which is about 2-3 times larger than 
for the static case. For larger switching times, the transport velocity decreases monotonously with $t_{\rm on}$.
To also compare the dispersion in the model and in the experiments, we show the standard deviation $\sigma$ also in Fig.~\ref{fig:exp}a, finding near quantitative agreement. 
It is instructive to visualize this also in a different way. To do this, 
we define the coefficient of variation as
$CV=\langle v\rangle /\sigma$ (quality factor), which is the ratio of the mean transport velocity over the
standard deviation, shown in Fig.~\ref{fig:exp}b. While there are notable deviations between experiments and simulations for small and for large $t_{\rm on}$ (Fig.~\ref{fig:exp}b), 
we find a rather good agreement in the regime $2s \leq t_{\rm on} \leq 8s$, where the transport is dominated by flashing, rather than by 
torque-saturation effects. This reflects that also in experiments the synchronization between particles and light-field leads to a 
transport with little dispersion.  
It is instructive to also resolve the time-evolution of the average particle speed in the $N$-th flashing periode, 
defined as 
$\langle \dot x \rangle(N)$, where the average is taken both over the particle ensemble and the $N$-th flashing period.
Here, both in experiments and in simulations, the transport converges to its steady state value within a few driving cycles 
in most cases, but takes significantly longer in those cases, where dispersion is large ($t_{\rm on}=1s$ and $t_{\rm on}=10s$).

\newpage 
\section{Conclusions}
\label{sec:conc}
Active colloids can synchronize with a sawtooth-shaped flashing-light field and self-organize into a pattern of coherently propagating density spikes. 
This pattern hardly disperses and yields a pulsating particle transport, which might be useful e.g. for targeted drug delivery applications.  
The transport velocity can be tailored by the parameters of the flashing motility field, and remarkably, it reverts when the particle's self-propulsion velocity exceeds a certain threshold.
Thus, the present setup can be used as a device for segregating particle ensembles. 
To observe the latter aspect also in experiments, it would be interesting to do 
experiments with mixtures of active colloids which are faster than the ones used in the present study, in the future. 
Our results can be straightforwardly generalized to more complex spatio-temporal motility patterns including random landscapes 
in space and time, and can be used as a platform to transfer ideas from 
potential ratchets \cite{Reimann1997,Reimann2002,Hanggi2009} to active systems. 

\section*{Acknowledgments}
H.L. and C.B. acknowledge financial support through the priority programme SPP 1726 of the Deutsche Forschungsgemeinschaft (DFG, German Research Foundation).
C.B. acknowledges financial support by the ERC Advanced Grant ASCIR (Grant No.693683).
B.t.H. gratefully acknowledges received funding
through a Postdoctoral Research Fellowship from the Deutsche
Forschungsgemeinschaft -- HA 8020/1-1.

\balance

\newpage
\bibliography{refs} 

\begin{mcitethebibliography}{46}
\bibitem[foo()]{footnote1}
Following Eq.~(\ref{unsaturated}), the orientations of particles in the steep
  $a$-segments change by a factor $b/a$ faster than in the $b$-segment. Since
  there are $a/b$ times less particles in the $a$-segments, the overall change
  in alignment for particles of any given initial orientation in the $a$- and
  in the $b$-segment is the same. More precisely, consider a representative
  initial ensemble with uniformly distributed positions and orientations, i.e.
  with distribution functions $P(x,\phi,0)=1/(2\pi L)$ for $x\in [0,L)$ and
  $P(\phi,0)=\int\limits_0^L {\rm d}x P(x,\phi,0)=1/(2\pi)$. Then, from the
  Smoluchowski equation corresponding to
  Eqs.~(\ref{tdyn},\ref{odyn},\ref{unsaturated}), after averaging over space,
  we obtain $P(\phi,dt)=P(\phi,0)$, where $dt$ represents an infinitesimal time
  step. Since we are not flashing, this argument can be repeated, yielding
  $P(\phi,t)=P(\phi,0)$, showing that no net alignment and hence no transport
  can occur at any time.\relax
\mciteBstWouldAddEndPunctfalse
\mciteSetBstMidEndSepPunct{\mcitedefaultmidpunct}
{}{\mcitedefaultseppunct}\relax
\EndOfBibitem
\end{mcitethebibliography}
\bibliographystyle{rsc} 

\end{document}